\documentclass[footinbib,aps,pra,reprint,superscriptaddress,nopacs]{revtex4-1}

\usepackage{graphicx}
\usepackage{amsmath,amsfonts,amssymb,bbm,bm}
\usepackage{verbatim}
\usepackage{braket}
\usepackage{mathtools}
\usepackage[absolute]{textpos}
\usepackage{ragged2e}

\begin{document}
\title{Double-pass multiple-plate continuum for high temporal contrast nonlinear pulse compression}

\author{Bo-Han Chen}
\author{Jia-Xuan Su}
\author{Jhan-Yu Guo}
\affiliation{Institute of Photonics Technologies, National Tsing Hua University, Hsinchu 30013, Taiwan}

\author{Kai Chen}
\affiliation{Robinson Research Institute, Faculty of Engineering, Victoria University of Wellington, Wellington 6012, New Zealand}
\affiliation{The Dodd-Walls Centre for Photonic and Quantum Technologies, Dunedin 9016, New Zealand}

\author{Shi-Wei Chu}
\affiliation{Department of Physics, National Taiwan University, Taipei 10617, Taiwan}
\affiliation{Molecular Imaging Center, National Taiwan University, 1, Sec 4, Roosevelt Rd., Taipei 10617, Taiwan}

\author{Hsuan-Hao Lu}
\affiliation{Elmore Family School of Electrical and Computer Engineering and Purdue Quantum Science and Engineering Institute, Purdue University, West Lafayette, Indiana 47907, USA}

\author{Chih-Hsuan Lu}
\email{lzch2000@hotmail.com}
\affiliation{Institute of Photonics Technologies, National Tsing Hua University, Hsinchu 30013, Taiwan}

\author{Shang-Da Yang}
\email{shangda@ee.nthu.edu.tw}
\affiliation{Institute of Photonics Technologies, National Tsing Hua University, Hsinchu 30013, Taiwan}
\date{\today}

\begin{abstract}
We propose a new architecture, double-pass multiple-plate continuum (DPMPC), for nonlinear pulse compression. In addition to smaller footprint, a double-pass configuration is designed to achieve substantial bandwidth broadening without incurring noticeable higher-order dispersion, thus improving the temporal contrast over those of traditional single-pass geometry when only quadratic spectral phase can be compensated. In our proof-of-concept experiment, 187~$\mu$J, 190-fs Yb-based laser pulse is compressed to 20~fs with high throughput (75~\%), high Strehl ratio (0.76) and excellent beam homogeneity by using DPMPC. Subsequently generated octave-spanning spectrum exhibits a significantly raised blue tail compared with that driven by pulses from a single-pass counterpart.
\end{abstract}

\maketitle
\textit{Introduction.---}Ti:Sapphire laser “dynasty” in ultrafast optics \cite{spence1991, wirth2011} has lasted three decades primarily because of the superb time resolution (down to 5~fs) \cite{ell2001} and high pulse energy (up to 250~J). However, the short upper-state lifetime (3.2 $\mu$s) and the requirement of green pump result in low repetition rate and low average power, which are unpleasant in applications such as material processing and coincident measurement \cite{ullrich2003}. Industrial grade Yb-doped lasers have received increasing attention as a scientific research tool in recent years. In addition to compactness and robustness, the gain media have sub-millisecond upper-state lifetime and can be directly pumped by diode to achieve high repetition rate (up to tens of MHz) and high average power (up to kilowatts). However, the relatively narrow gain bandwidth only supports sub-picosecond to picoseconds pulses, leaving the applications based on few-cycle pulses still the playground of Ti:Sapphire laser \cite{jeong2018}.

\begin{figure*}[bt!]
\centering
\includegraphics[width=0.6\textwidth]{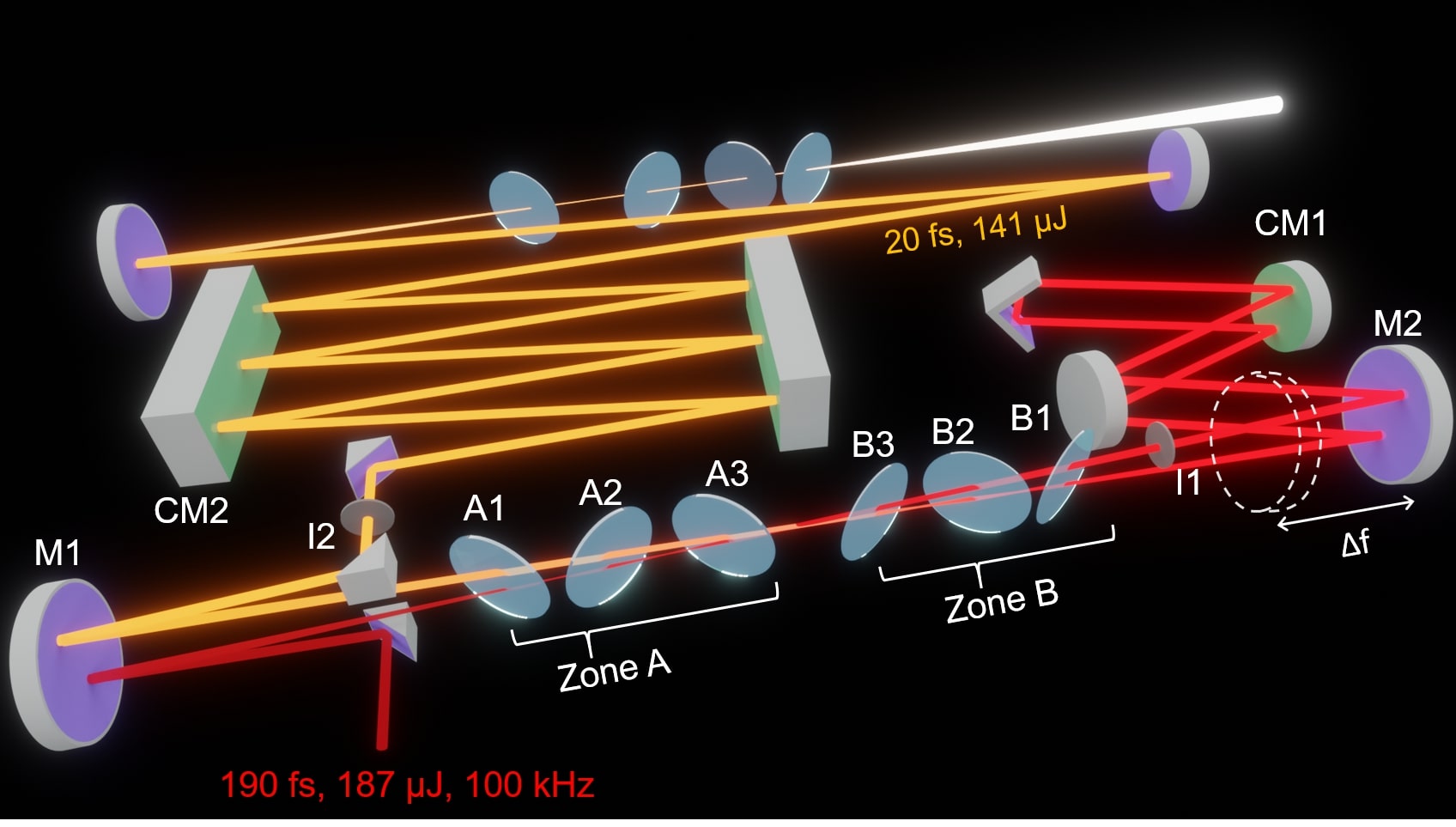}
\caption{Experimental setup of DPMPC. M\#: concave mirrors; CM\#: chirped mirror sets; I\#: irises, A\# and B\#: quartz plates.}
\label{fig:001}
\end{figure*}

A straightforward solution to improve the pulse width of industrial grade Yb-doped lasers is using a nonlinear pulse compressor. Various nonlinear pulse compression techniques, such as hollow-core fiber (HCF) \cite{jeong2018, nagy2019}, multi-pass cell \cite{schulte2016, lavenu2018, ueffing2018} and multiple-plate continuum (MPC) \cite{lu2014, beetar2018, lu2019}, have been explored and applied to metrology, time-resolved spectroscopy \cite{tamming2021}, generation of coherent extreme ultraviolet \cite{huang2018} and terahertz radiation \cite{kanda2021}. HCF compressors can deliver exceptionally high pulse energy and beam quality, but are extremely sensitive to beam pointing fluctuation. Both multi-pass cell and MPC implement spectral broadening in a cascaded fashion. Multi-pass cell accumulates a small nonlinear phase shift $\phi_{NL}$ per pass, while the intensity is maintained by (chirped or unchirped) focusing mirrors. The overall spectral broadening (depending on the total $\phi_{NL}$) can be large after many passes, and the resulting spectral shape is highly symmetric due to the distributed nature of nonlinear effect. However, multi-pass cell requires special mirrors with specific group-delay dispersion (GDD) over octave-spanning bandwidth to achieve sub-10~fs pulses \cite{balla2020}. Besides, the lengthy optical path caused by multiple bounces in the cell could be a trouble for temporal synchronization in pump-probe applications. In contrast, traditional MPC pushes $\phi_{NL}$ close to the limit in each plate, while the risk of self-focusing-induced damage or wavefront distortion can be managed by adjusting the thicknesses of and distances between the plates. Single-cycle pulse can be generated in a simple optical setup without demanding alignment \cite{lu2014}. The downside of MPC is the poorer temporal pulse contrast due to asymmetric spectral shape (when self-steepening matters) coupled with stronger higher-order dispersion. Customized chirped mirrors are often required to effectively remove the nonlinear chirp \cite{lu2019}.

In this contribution, we demonstrate double-pass multiple-plate continuum (DPMPC), aiming to improve the temporal contrast of few-cycle MPC pulses in a compact setup. In this architecture, the moderately broadened spectrum and predominantly quadratic spectral phase after the first pass permit nearly perfect dispersion compensation by using off-the-shelf chirped mirrors with constant GDD. The “well-behaved” output pulse resumes a self-phase modulation (SPM)-dominated spectral broadening in the second pass with reduced higher-order dispersion. After removal of linear chirp and the introduction of an additional MPC stage, we achieve octave-spanning spectrum with a blue tail enhanced by $\sim$10~dB.

\textit{Experimental results.---}
Figure~\ref{fig:001} shows the experimental setup of DPMPC. A Yb:KGW laser (Carbide, Light Conversion) delivers 190~fs, 187~$\mu$J pulses at 100 kHz repetition rate and 1030~nm central wavelength. A concave mirror M1 (f = 50~cm) focuses the beam onto a 200-$\mu$m-thick quartz plate to initiate the spectral broadening process, where the peak intensity is estimated to be 6.52~TW/cm$^{2}$. Two subsequent quartz plates with the same thickness are inserted behind the self-focused position from the previous plate, such that maximal broadening can be achieved while no damage is observed \cite{lu2014}. These three plates form “Zone A”, after which no spectral broadening is observed by adding additional plates [dashed, Fig. 2 (a)].

\begin{figure}[b!]
\centering
\includegraphics[width=3.3in]{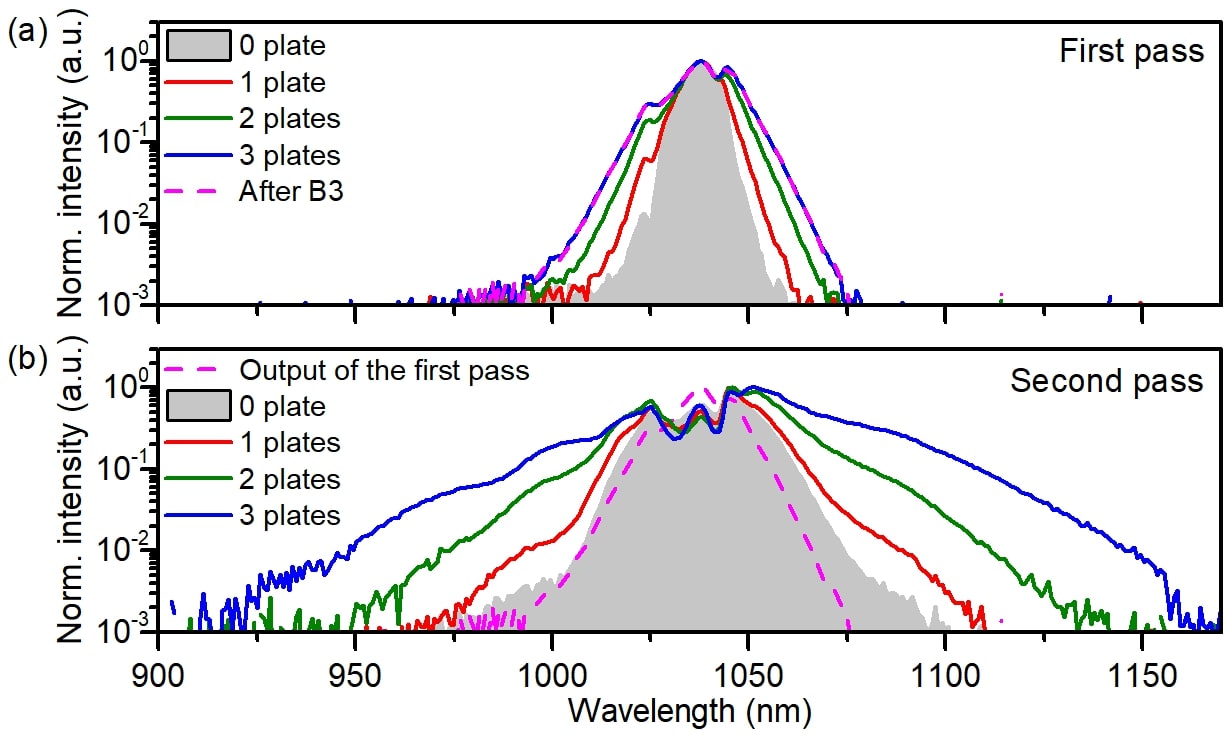}
\caption{Spectral evolution during the (a) first pass, and (b) second pass in DPMPC.}
\label{fig:002}
\end{figure}

After the first pass, an iris I1 is used to improve the beam homogeneity by blocking the rim with less spectral broadening. The pulse energy becomes 173~$\mu$J, corresponding to 92.5~\% throughput. A second concave mirror M2 (f = 50~cm) is used to collect the diverging beam after Zone A. Note that the position of M2 is deliberately detuned from the cavity condition (by $\Delta$f = 20~mm) such that the beam is slightly converging instead of being collimated. A pair of chirped mirrors CM1 and a retroreflector are employed to compensate the dispersion accumulated during the first optical pass (including all the six plates) and redirect the beam back to M2 at the same angle but at a lower position. The dechirped pulse is characterized by polarization-gating frequency-resolved optical gating (PG-FROG). After 4 bounces of chirped mirrors (-2000~fs$^{2}$ in total), the pulse width (FWHM) is compressed to 54.2~fs [blue solid, Fig. 3(a)] with 164~$\mu$J energy and a Strehl ratio (peak power relative to that of a Fourier-limited pulse of the same energy) of 0.82. As the beam is slightly converging, the focal spot is moved closer to M2, where the peak intensity increases to $\sim$20.1~TW/cm$^{2}$. In order to avoid multiple filaments and optical breakdown, thinner (50~$\mu$m) quartz plates are used in Zone B. The plates are put in the same way as in Zone A, and the output beam is collected by M1, spatially filtered by I2, and guided to a second set of chirped mirrors CM2 by a pick-off mirror. After 8 bounces of chirped mirrors (-480~fs$^{2}$ in total), pulse width is down to 20~fs [blue solid, Fig. 3(c)] with 141~$\mu$J energy (corresponding to throughputs of 89.5~\% and 75.4~\% for the second pass and the entire DPMPC system) and a Strehl ratio of 0.76, respectively.

\begin{figure}[bt!]
\centering
\includegraphics[width=3.3in]{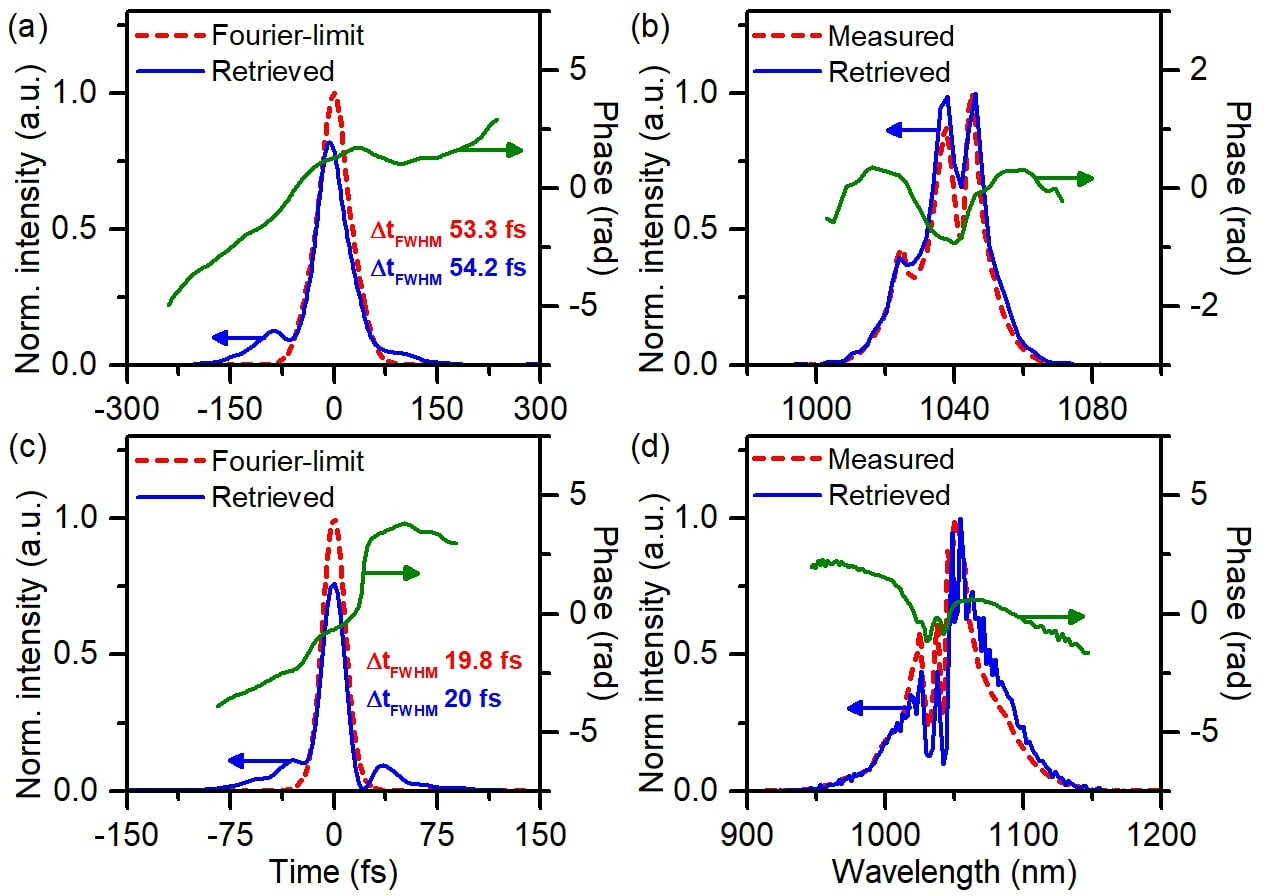}
\caption{PG-FROG measurement of pulses after the (a,b) first, and (c,d) second passes in (a,c) time, and (b,d) frequency domains, respectively. The retrieved intensity profiles in (a) and (c) are normalized to the energy of their Fourier-limit cases. FROG error: 0.36~\% for (a, b) and 0.51~\% for (c, d).}
\label{fig:003}
\end{figure}

Spectral evolutions during the two passes (Fig.~\ref{fig:002}) exhibit signatures of SPM, where the relatively symmetric shapes support Fourier-limit pulses of higher temporal contrast [red dashed, Figs ~\ref{fig:003}(a) and ~\ref{fig:003}(c)] than those arising from self-steepening or white-light generation \cite{bradler2009}. Figure~\ref{fig:002}(a) also confirms that plates in Zone B do not disturb spectral broadening in the first pass, as the spectrum after plate B3 (dashed) resembles that after plate A3 (blue solid). Limited by the folded geometry, spectra in the second pass [Fig.~\ref{fig:002}(b)] are measured after iris I2. Given the input spectrum (grey shaded) is slightly broader than the
output spectrum of the first pass [blue, Fig.~\ref{fig:002}(a)], plates in Zone A could have minor impact on spectral broadening in the second pass while no noticeable extra loss or beam deterioration is observed.

\begin{figure}[b!]
\centering
\includegraphics[width=3.3in]{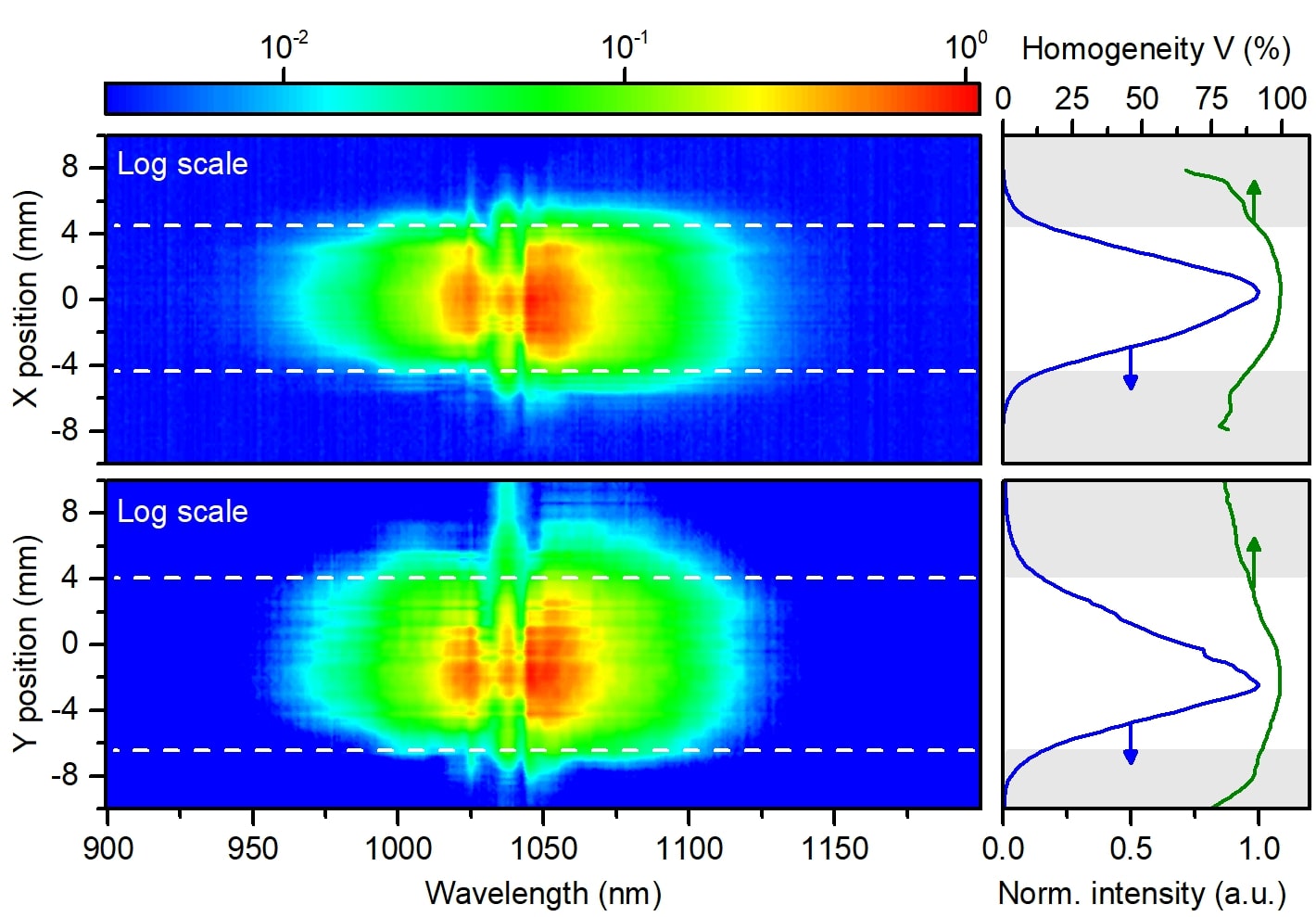}
\caption{Homogeneity measurement along x- and y-axis of the DPMPC beam. The white-dashed lines on the left panel define the borders where light intensity drops to 1/e$^{2}$ of its peak.}
\label{fig:004}
\end{figure}

Figure~\ref{fig:004} shows the spatially resolved power spectrum of DPMPC beam measured by acquiring the power spectrum after a 600-$\mu$m-wide slit scanning across the beam cross section. The 2D patterns exhibit virtually no tilted structure and the calculated homogeneity values V \cite{weitenberg2017} are higher than 87~\% within the 1/e$^{2}$ diameter for both x- and y-axis, indicating good beam homogeneity. Beam caustic measurement (at 1064 nm) also shows great beam quality with $M_X^2$ and $M_Y^2$ values 1.08 and 1.16, respectively.

\begin{figure}[b!]
\centering
\includegraphics[width=3.3in]{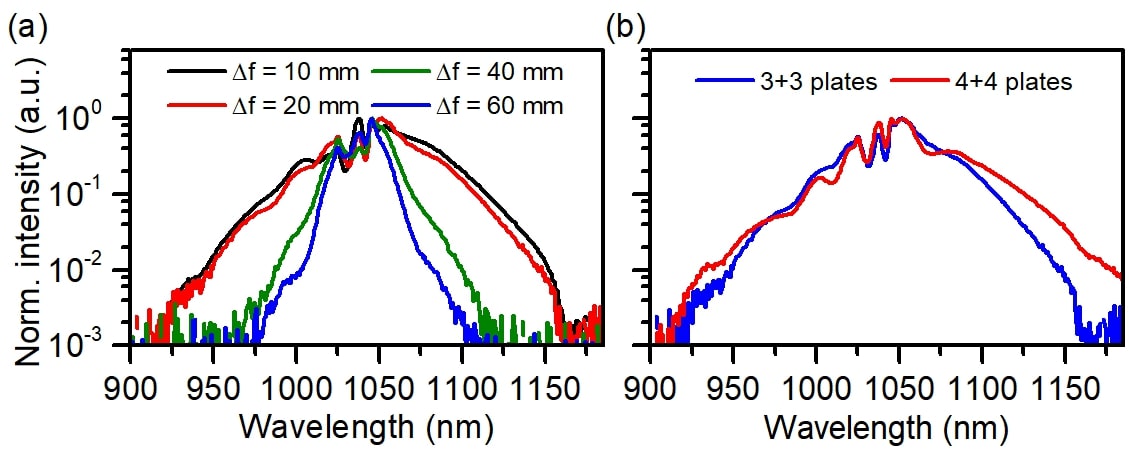}
\caption{Output spectra of DPMPC measured at different (a) M2 positions, and (b) numbers of quartz plates.}
\label{fig:005}
\end{figure}

The flexibility of DPMPC is experimentally verified in two aspects. First, output spectra are measured when the concave mirror M2 is deviated from the cavity condition by different amounts of $\Delta$f [Fig.~\ref{fig:005}(a)]. For $\Delta f <10$~mm, the back focal plane (where plate B1 is supposed to be placed) will be too close to Zone A, leaving insufficient space for Zone B. As $\Delta$f increases, output spectrum is narrower because of the bigger focal spot size (thus lower peak intensity). Spectral broadening in the second pass would be nominal if $\Delta f > 60$~mm. This indicates that DPMPC can operate over a wide range of parameters without failure. Here we elect to operate at $\Delta$f = 20~cm, a condition which leaves us enough space for DPMPC while achieving near-optimal spectral broadening [red, Fig.~\ref{fig:005}(a)]. Secondly, we tried DPMPC with different amounts of plates (3+3 plates and 4+4 plates) at fixed $\Delta$f = 20~mm [Fig.~\ref{fig:005}(b)]. As expected, the spectrum arising from 4+4 plates (red) is broader than the case of 3+3 plates (blue) owning to its longer interaction length (thus more total $\phi_{NL}$). However, the resulting spectrum slightly exceeds the high reflection coating range of the optics (M1), causing additional power loss after compression. Furthermore, the additional dispersion from the material requires more amount of GDD and even third-order dispersion (TOD) engineering for quality dispersion compression. As a result, we choose 3+3 plates for most of our demonstrations.

\begin{figure}[t!]
\centering
\includegraphics[width=3.3in]{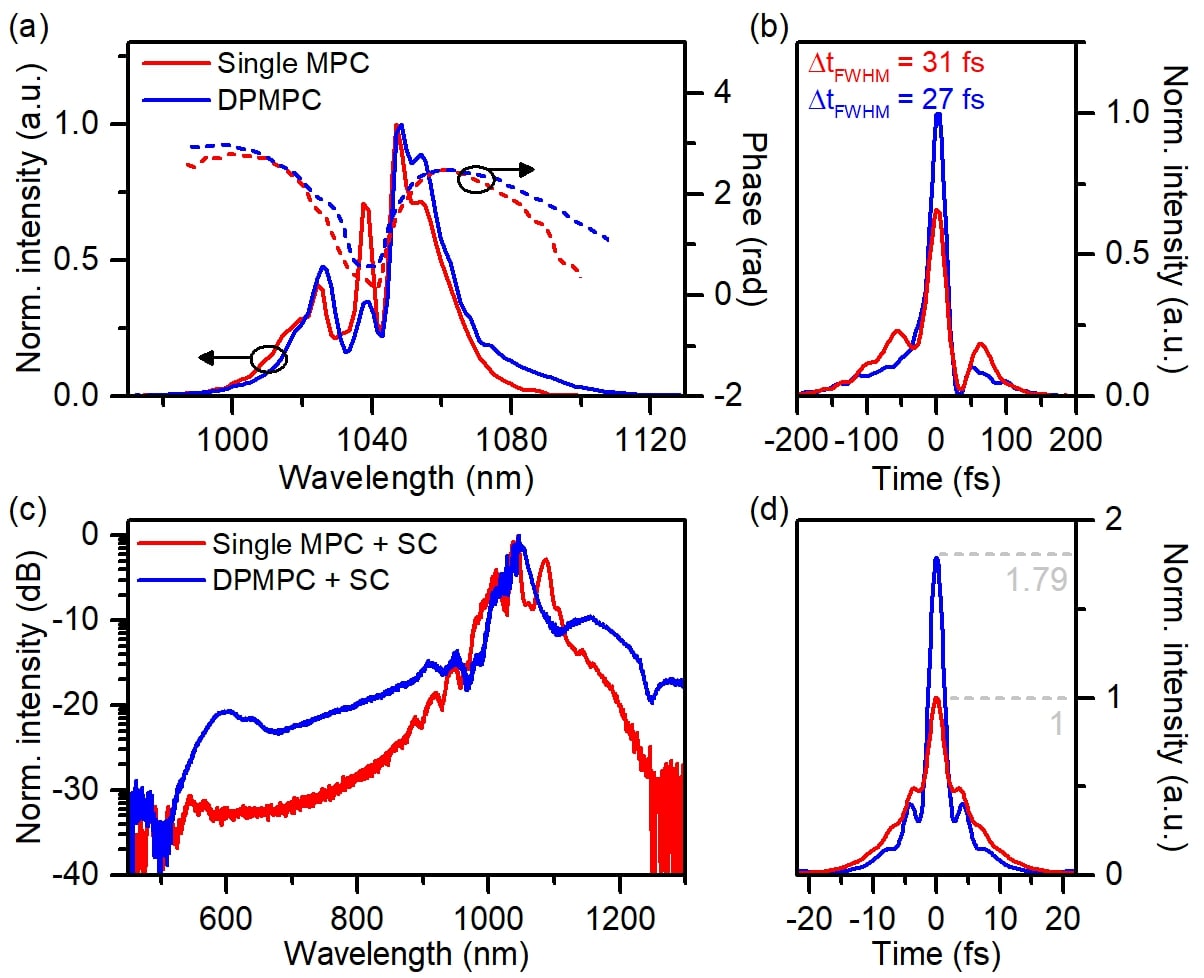}
\caption{(a, b) Pulses from single-pass MPC (red) and DPMPC (blue) retrieved by PG-FROG, where the areas under curves in (b) are proportional to the pulse energies. (c) The corresponding SC spectra generated in an extra MPC stage. (d) The Fourier-limit temporal intensities (normalized to the same pulse energy) corresponding to the spectra in (c).}
\label{fig:006}
\end{figure}

The advantage of DPMPC over traditional single-pass geometry in terms of subsequent octave-spanning spectrum generation is demonstrated in Fig. 6. With the same input pulse (190~fs, 187~$\mu$J) and using six pieces of plates, both techniques achieve supercontinuum (SC) [550-1300~nm, Fig.~\ref{fig:006}(c)] in two stages (DPMPC + MPC and MPC + MPC). Since one single-pass MPC only manages to compress the input pulse to 31~fs [red, Fig.~\ref{fig:006}(b)], the DPMPC is deliberately tuned to provide a similar pulse width [blue, Fig.~\ref{fig:006}(b)] for fair comparison. The retrieved spectral phase curves [dashed, Fig.~\ref{fig:006}(a)] show predominant quartic dependence, where the fitted
fourth-order dispersion coefficient $\beta_4$ of single-pass MPC ($2.7 \times 10^{6}$~fs$^{4}$) is 5 times larger than that of DPMPC ($5.4 \times 10^{5}$~fs$^{4}$). The smaller residual $\beta_4$ of DPMPC supports a smoother pulse shape and better temporal contrast [blue, Fig.~\ref{fig:006}(b)], which are favorable for the subsequent SC generation. As shown in Fig.~\ref{fig:006}(c), the blue tail is enhanced by up to $\sim$10~dB (at 600 nm). The more-symmetric spectral shape substantially suppresses the side peak level of Fourier-limited pulse shape from 49~\% to 22~\% [Fig.~\ref{fig:006}(d)], equivalent to peak power enhancement by a factor of $\sim$1.79 under the same pulse energy.

\textit{Conclusion.---}In summary, we propose and demonstrate DPMPC scheme to compress 100 kHz, 190~fs, 187~$\mu$J pulses at 1030~nm down to 20~fs with high throughput (75~\%), high Strehl ratio (0.76), and good spatio-spectral homogeneity. The results show that an intermediate compression stage provides a well-behaved pulse that can resume a SPM-dominated spectral broadening with less accumulation of higher-order dispersion. Pulses of high temporal contrast can arise after dechirping by only using off-the-shelf chirp mirrors with constant GDD. Subsequent SC generation shows that the blue spectral tail is enhanced by up to 10 dB. The enhanced spectral density in the visible makes DPMPC attractive in spectroscopic investigation of photovoltaic materials.

\begin{acknowledgments}
This work was supported by the Brain Research Center under the Higher Education Sprout Project, co-funded by the Ministry of Education and the Ministry of Science and Technology in Taiwan. Funding was provided by the Ministry of Science and Technology of Taiwan (MOST 109-2112-M-007-013, MOST 109-2811-M-007-517, MOST 110-2811-M-007-508, MOST 110-2321-B-002-012), and the BRC research center funding (110-2634-F-007-025).
\end{acknowledgments}

\providecommand{\noopsort}[1]{}\providecommand{\singleletter}[1]{#1}%

\end{document}